\documentclass[
     final            
  ]
  {aipproc}
\usepackage{graphicx}
\usepackage{color}
\usepackage{dcolumn}
\usepackage{bm}
\layoutstyle{8x11single}

\begin{document}

\newcommand{\oldbibitem}[2][]{}	
\let\oldbibitem\bibitem		
\newcounter{totalbibitem}
\renewcommand{\bibitem}[2][]{\oldbibitem[#1]{#2}\addtocounter{totalbibitem}{1}}

\title{Bi-functional nonlinearities in monodisperse ZnO nano-grains --
  Self-consistent transport and random lasing}

\classification{42.25.Dd, 42.55.Zz, 72.15.Rn, 78.20.Bh}
\keywords{disordered systems, transport theory, random lasing, non-equilibrium}

\author{Andreas Lubatsch$^1$ and Regine Frank$^{2*}$ \footnote{Phone:  +49-7071 / 29-73434; email:
  r.frank@uni-tuebingen.de; web: www.uni-tuebingen.de/photonics}}{
  address={$^1$ Georg-Simon-Ohm University of Applied Sciences,
Ke{\ss}lerplatz 12, 90489 N\"urnberg, Germany,\\
$^2$ Institute for Theoretical Physics, Optics and Photonics,
  Eberhard-Karls-Universit\"at, Auf der Morgenstelle 14, 
  72076 T\"ubingen, Germany}
}

\begin{abstract}

We report a quantum field theoretical description of light transport and
random lasing. The Bethe-Salpeter equation is solved including maximally
crossed diagrams and non-elastic scattering. This is the first theoretical
framework that combines so called off-shell scattering and lasing in random media.
We present results for the self-consistent scattering mean free path that varies
over the width of the sample. Further we discuss the density dependent correlation
length of self-consistent transport in disordered media composed of
semi-conductor Mie scatterers.  

\end{abstract}

\maketitle


\section{Introduction}
  Random lasers and their coherence properties are recently investigated
  theoretically as well as experimentally \cite{Stone,Jacquod,Redding1,Bell}.
However efficient theoretical methods that may treat strongly scattering solid
state random lasers, including non-linear gain and gain saturation, are still of
urgent need. One ansatz to reach this goal is to employ methods from quantum
field theory that have proven to be efficient in solving strong
localization of photons in random \cite{Tiggelen1,Tiggelen2} and complex
media \cite{Frank2011,NPHOT}. In this article we investigate the spatial
  coherence properties of different random laser samples theoretically. The
  samples only vary in their filling with spherical ZnO Mie scatterers. 
  Besides the coherence within these systems we discuss the self-consistent scattering mean free
path $l_s$ of random lasers. We show that the scattering mean free path
$l_s$ of random lasers is not only a material characteristic and dependent to
the filling as it has been often estimated in literature
\cite{Akkermans,Chang,Kalt09}. Instead it changes in depth of the
  sample and therefore depends on the nonlinear self-consistent gain in strongly
  scattering solid-state random lasers, especially at the surface.\\

\section{Model}

The theoretical model is based on an extended approach of the Bethe-Salpeter
equation including maximally crossed diagrams. Additionally we model the
scattering nano-grains by means of ZnO semi-conductor Mie spheres. This
implies so-called off-shell scattering which is an implicit characteristic of a
complex refractive indexed medium \cite{Sakurai}. Consequently it leads to
a renormalized condition for local energy conservation, the Ward identity for
active Mie spheres \cite{Lubatsch}. We discuss how this approach
can be expanded to a more sophisticated frame using non-equilibrium Keldysh
theory in order to cover properly for locally occurring electromagnetically
induced transparency (EIT).

 The system we consider consists of a randomly scattering
medium in the form of a slab geometry \cite{Wiersma}. This slab is
finite $d$ sized in the $z$-dimension and assumed of to be
of infinite extension in the $(x,y)$ plane (see Fig.(\ref{fig01})). In experimentally
relevant situations this refers to film structures of thickness 
up to 32 $\mu m$. 
The spherical Mie scatterers \cite{Mie,MoskMie} are embedded in a homogeneous host material which
is considered to be passive. Both, scatterer 
and host medium, are described by means of a complex dielectric function 
$\epsilon_s$ and $\epsilon_b$, respectively. The scatterers are 
modeled to be optically active ZnO, with a refractive index for the passive
case of $2.1$. The imaginary part of the permittivity $\rm Im\, \epsilon_s$, comprised in the gain is
self-consistently derived.
This sample is optically pumped in order to achieve a
sufficient electronic population inversion within the active 
medium of the scatterers by means of an incident pump laser, 
perpendicular onto the $(x,y)$-surface of the random laser.
The laser feedback is guaranteed by multiple scattering. The same mechanism
actually supports stimulated emission and hence coherent light intensity
within the setup. The so generated laser intensity then may leave the
sample through both open surfaces of the sample geometry, the dissipation
channels. The emitted light is eventually observed at the surface of the
sample in the form of lasing spots which comprise to a lasing mode.
These lasing modes are of a characteristic size which is determined through
this theory to be the correlation length $\xi$ within the mass term of the
diffusion pole. The latter depends only on system 
parameters such as scatterer size, wavelength of the pump source, filling
fraction and film thickness.

\begin{figure}
\begin{minipage}[hbt]{0,4\textwidth}
		\includegraphics[clip,scale=0.4]{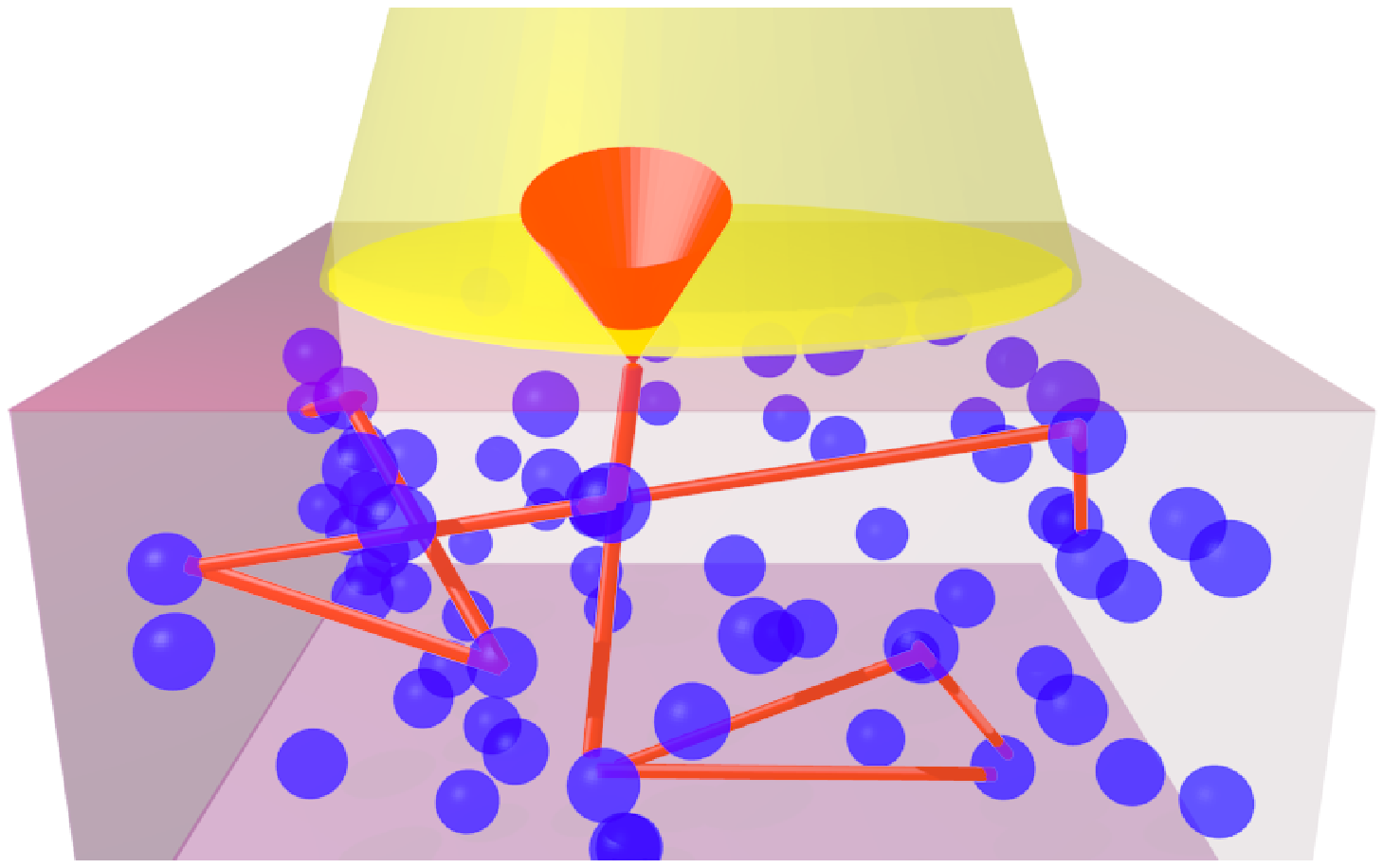}
\end{minipage}
\begin{minipage}[hbt]{0,6\textwidth}
		\includegraphics[clip,scale=0.4]{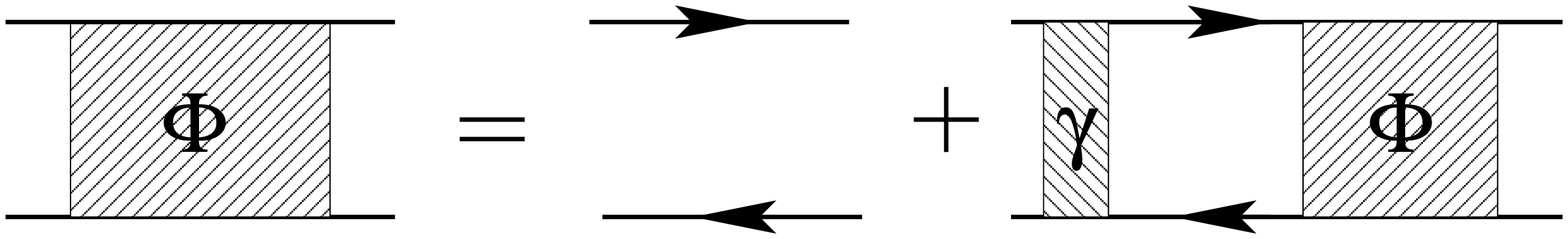}\label{fig01}
		\caption{(a) ZnO spherical Mie scatterers at random locations (blue) are optically
    pumped from above (wide yellow beam). The pumping yields an inversion
    of the atomic occupation number within the ZnO causing 
    stimulated emission of light (orange light paths). The emitted intensity
    multiply scatters and concentrates due to the samples density
    distributions. At the laser threshold the system experiences a phase
    transition and second order coherent intensity, meaning coherence in space
    and time, escapes the system through
    its surfaces (orange cone). The scatterers radius is $r_0 = 600 nm$,
    $\lambda= 723 nm$ and the samples finite dimension is of $d=32 \mu m$. (b) Diagrammatic representation of the Bethe-Salpeter equation. The four-point
correlator $\Phi$ (on the left) is given as self-consistent integral over $k$ and $k'$ relating the Green's
functions and the irreducible vertex $\gamma$ including multiple scattering as
well as all interference effects of time reversal processes (on the right).}
		\label{fig01}
\end{minipage}
\end{figure}

The field-field correlation or coherence length 
of the propagating lasering intensity, is derived by means 
of the field theoretical approach of localization of photons based on the 
theory  by Vollhardt and W\"olfe \cite{Woelfle80}. Non-equilibrium band-structure calculations with a basic Hubbard
model \cite{Frank12} for Zno bulk prove the existence density of states in the
semi-conductor gap and optical
gain for the non-equilibrium situation. The
latter indicates electrical/optical induced transparency (EIT)
processes for high energy pumping of solid state random
lasers. Consequently no optical gap is observed. It is washed
    out due to uncorrelated disorder (see next section) and the mentioned
    non-equilibrium processes. In addition lasing
    especially in the semi-conductor gap may be observed, where non-equilibrium is
    not necessarily to be assumed in the laser model, however the stationary
    state defines the threshold of the laser.

\section{Transport Theory}

The equation of motion for the electric field of stimulated emitted
light $\Psi_\omega(\vec{r}\,)$ within the sample is given by the wave equation  

\begin{equation}
 \label{Wave}
\frac{\omega^2}{c^2} \, \epsilon (\vec{r}\,) \Psi_\omega(\vec{r}\,) 
+ \nabla ^2 \Psi_\omega( \vec{r}\,) 
= -i \omega \frac{4\pi}{c^2}  j_\omega(\vec{r}\,)\ , 
\end{equation} 

where we denote $c$ to be the vacuum speed of light and  $j_\omega (\vec{r}\,)$ 
the external source. The dielectric constant  
$ \epsilon(\vec{r}\,) = \epsilon_b + \Delta\epsilon\, V(\vec{r}\,)$,
where  the dielectric contrast has been defined according to 
$\Delta\epsilon = \epsilon_s - \epsilon_b$, including a random arrangement of scatterers in terms of  
$V(\vec{r}\,) = \sum_{\vec{R}} S_{\vec{R}}\,(\vec{r}-\vec{R}\,)$, with 
$S_{\vec{R}}\,(\vec{r}\,)$ a localized shape function 
at random locations $\vec{R}$. The intensity is then related to the 
field-field-correlation function $\Phi$, often referred to as the 
four-point-correlation, 
$\Phi = \langle \Psi(\vec{r},\,t\,) \Psi^*(\vec{r}\,',t\,'\,)\rangle$. 
Here, the angular brackets $\langle \ldots \rangle$ refer to the disorder average
or ensemble average of this random system \cite{PingSheng,Jalickee}. In order to calculate the  
field-field-correlation $\Phi$ the Green's function formalism 
is best suited. The wavefunction of the electromagnetic field reads

\begin{equation} 
 \Psi(\vec{r},\,t\,) = 
\int {\rm d}^3r\,'  
\int {\rm d}t\,'   
G( \vec{r} \, ,\vec{r} \,'\, ;\,t\,,t'\,)      j(\vec{r}\,'\,,t'\,)\,.
\end{equation}

The  single-particle Green's function Eq. (\ref{SP_Green_function_field}) is related 
to the (scalar) electrical field, by inverting the (non-linear) wave equation
Eq. (\ref{Wave}) . It reads in in the density
approximation of independent scatterers \cite{Tiggelen1, Frank2011}
\begin{equation} 
\label{SP_Green_function_field} 
G(\omega,\vec{q})=\frac{1}{\epsilon_b(\omega/c)^2  -|\vec q|^2 - \Sigma^\omega_{\vec q}}
\end{equation} 
where $\omega$ is the light frequency
    and $\epsilon_b$ is the dielectric function of the space in between the
    scatterers, $\vec q$ is the wave vector. $\Sigma^\omega_{\vec q}=n\cdot T$. $T$ is the complex valued T-Matrix of the
single scatterer, $n$ is the volume filling fraction and  $\Sigma^\omega_{\vec q}$ is the single particle
self-energy including Mie scattering of the spheres coupled the non-linear
response of the amplifying material. The scatterers are bi-functional in the
sense that the semi-conductor structure amplifys light
    by generating light matter bound states yielding gain which renormalizes
    the resonance and leads to gain saturation. This behavior is typical for
  strongly scattering solid state random lasers comprised of pure semi-conductor powder and in theory it goes far
  beyond previously existing approaches, e.g. \cite{Stone,Jacquod}.

In order to study transport in the above introduced 
field-field-correlation we consider the  4-point correlation function, 
defined now in terms of the non-averaged Green's functions, i.e. the retarded  
$G^R$ and the  advanced Green's function $G^A$, where now we find 
$\Phi \sim  \langle G^R G^A\rangle$. The intensity correlation obeys an equation of motion itself, the
Bethe-Salpeter equation (BS) \cite{Frank2011}, given in coordinate space given as

\begin{eqnarray}
\label{BS_real_space_neu_01}
\Phi (r_1,r_1';r_2,r_2')
=
G^R (r_1,r_1')  G^A (r_2,r_2') + \sum_{r_3,r_4,r_5,r_6} G^R (r_1,r_5)  G^A
(r_2,r_6) \gamma  (r_5,r_3;r_6,r_4) \Phi  (r_3,r_1';r_4,r_2').
\end{eqnarray}

\begin{figure}[t]
  \includegraphics[clip,scale=0.35]{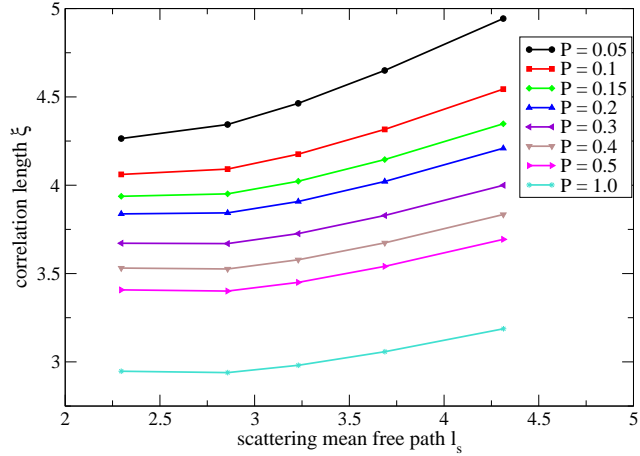}
  \caption{
    Calculated correlation length $\xi$ of the random lasing modes as a
    function of the calculated scattering mean free path $l_s$. Both length
    scales are given in units of the scatterer radius $r_0$. Different curves 
    correspond to different strengths of the pump intensity P, given in 
    units of transition rate $\gamma_{21}$.
    The different points 
    along a given curve correspond to different filling fractions of the 
    zinc oxide scatterers. The symbols from left to right correspond to
    filling fractions of $60\%$, $50\%$, $45\%$, $40\%$, and $35\%$. 
    }
    \label{corr1}
\end{figure}

In the BS, we introduced the irreducible vertex function
$\gamma  (r_5,r_3;r_6,r_4)$ which represents all scattering 
interactions inside the disordered
medium of finite size. The irreducible vertex is discussed in the
given references in detail but we mention here that beyond ladder diagrams ({\it
Diffuson}) so
called maximally crossed diagrams ({\it Cooperons}) are included. This is
actually a matter of course within the self-consistent theory of localization
but it exceeds the usual description of the  Bethe-Salpeter equation. Local controlled energy non-conservation is
incorporated by means of the Ward identity \cite{Lubatsch}. To account for the particular form of the system geometry, Wigner
coordinates are chosen, where a full Fourier transform of the spatial
coordinates within the infinite extension of the  $(x,y)$-plane is used. We use relative $\vec{q}_{||}=(q_x,q_y)$ and center-of-mass 
momentum $\vec{Q}_{||}=(Q_x,Q_y)$  variables. However, the finite $z$-coordinate 
of the slab is transformed into relative and center-of-mass real-space 
coordinates, i.e. $z$ and $Z$ respectively. In this 
representation only the relative coordinate is Fourier transformed.
This procedure is justified because the relative coordinates of the 
intensity correlation are related to the scale of the oscillating electric
field, whereas the center-of-mass coordinates are related to the scale 
of the collective behavior of intensity, which is a significantly 
larger scale. Given that the thickness of the slab is much larger than 
the wavelength of the laser light as discussed above, a Fourier transform 
with respect to this perpendicular relative coordinate is perfectly
acceptable. In this representation the BS equation, Eq. (\ref{BS_real_space_neu_01}), 
may be rewritten according to 

\begin{eqnarray}
\!\!\!\!\!\!\!\!\!\!\!\!\!\!&&\Bigg[
\Delta \Sigma
+
2 {\rm Re\,} \epsilon \omega\Omega
-
\Delta \epsilon\omega^2
-
2\vec{p}_{||}\cdot \vec{Q}_{||}
+2ip_z\partial_Z
\Bigg] \Phi_{pp'}^{Q_{||}} (Z  , Z') \\
&&=      \label{TT_kinetic} 
\Delta G_p (Q_{||}; Z, Z') 
\delta(p-p') + 
\sum_{Z_{34}}
\Delta G_p (Q_{||})
\int \!\!\!\frac{{\rm d}p'' }{(2\pi)^3}
\gamma_{pp''}^{Q_{||}}  (Z,Z_{34})
\Phi_{p''p'}^{Q_{||}}  (Z_{34} , Z')
\nonumber
\end{eqnarray}

where we used the abbreviation $\Delta G \equiv G^R - G^A$.
The rewritten BS equation, Eq. (\ref{TT_kinetic}), also known as
kinetic equation, therefore is seen to be a differential equation in 
finite center-of-mass coordinate $Z$ along the limited dimension of the slab.
This differential equation is accompanied by suitable boundary 
conditions accounting for the reflectivity of the sample surfaces. 
Eq. (\ref{TT_kinetic}) is solved in terms of an 
expansion of the correlation $\Phi$ into its moments, identified as
energy density and energy current density correlation, respectively. A self-consistent expression for the diffusion constant is derived, 
accompanied by a pole structure within the energy density expression.

\begin{eqnarray}
\Phi_{\epsilon\epsilon}(Q,\Omega)=
\frac
{N_{\omega}(Y)}
{
\Omega + i D Q^2 -i D \xi^{-2} }.
\label{Pole}
\end{eqnarray}

The last term in the denominator $-i D \xi^{-2}$ is the so called mass term
which is present for all kinds of complex media and off-shell scattering.
 
\begin{figure}[t]
  \includegraphics[clip,scale=0.35]{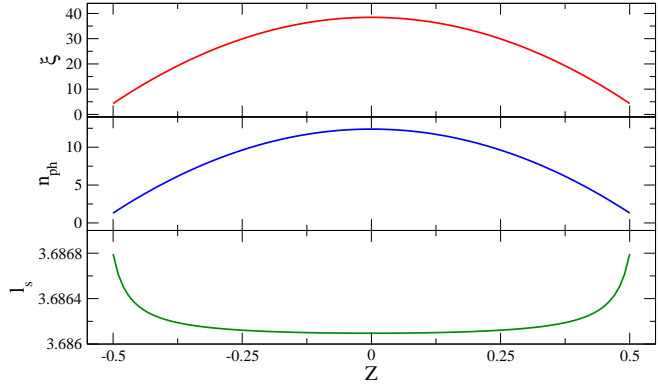}%
  \caption{
    Calculated correlation length $\xi$, photon number density $n_{ph}$ and
    scattering mean free path $l_s$ across the slab geometry from 
    surface to surface. The parameter set is a filling fraction of $40\%$
    and a pump rate of $P=0.1\gamma_{21}$. 
    }
    \label{corr2}
\end{figure}

\section{T-Matrix and Lasing}

The scattering properties of the disordered sample are included by means of an
independent scatterer approximation. $\Sigma= n\cdot T$, where T is the
complex valued T-Matrix of the single Mie sphere \cite{Lubatsch}, necessarily
this is ``off-shell''. The
conservation laws are represented by the Ward identity. The incorporation of
the lasing properties go by far beyond that approach. Even though the
equations look uncomplicated, the numerical efforts for convergency of the 3-dimensional system are non-trivial. The lasing behavior in terms of the atomic occupation 
number is described by means of the following four-level laser rate
equations \cite{Siegmann}

\begin{eqnarray} 
\frac{\partial N_3}{\partial t}  
&=&  
\frac{N_0}{\tau_{P}}  - \frac{ N_3}{\tau_{32}} \\ 
\frac{\partial N_2}{\partial t}  
&=&   
\frac{N_3}{\tau_{32}}  -   \left(\frac{1}{\tau_{21}} 
+ \frac{1}{\tau_{nr}}\right)N_2 -   
\frac{\left( N_2 -N_1\right)}{\tau_{21}} n_{ph} \\ 
\frac{\partial N_1}{\partial t}  
&=&     
\left(\frac{1}{\tau_{21}}+ \frac{1}{\tau_{nr}}\right)N_2   
+ \frac{\left( N_2 -N_1\right)}{\tau_{21}} n_{ph}  
-   \frac{ N_1}{\tau_{10}} \label{third_LRG} \\ 
\frac{\partial N_0}{\partial t}  
&=&    
\frac{N_1}{\tau_{10}} - \frac{N_0}{\tau_{P}} 
\end{eqnarray}  

\begin{figure}[t]
  \includegraphics[clip,scale=0.35]{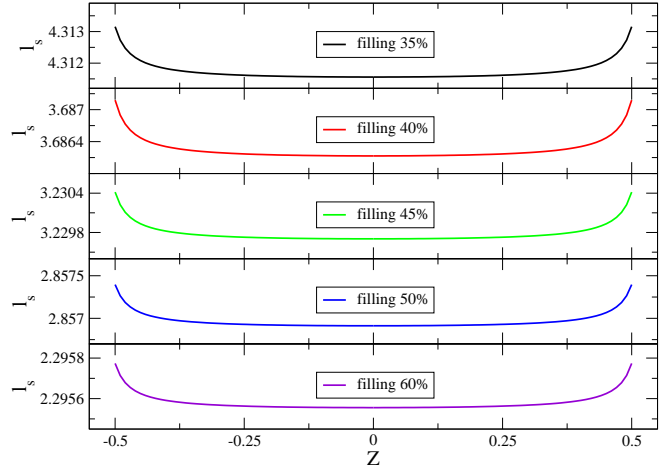}
  \caption{
    Calculated scattering mean free path $l_s$ across the depth $Z$ of the
    random lasing film for various filling fractions as indicated in
    the legends. The  pump rate is set to be  $P=0.5\gamma_{21}$.
    $l_s$ dependends on the atomic inversion, i.e. on the depth dependent imaginary
    part of the dielectric function ${\rm Im}\,\epsilon_s$, which has been
self-consistently
    calculated.
    }
    \label{fig04}
\end{figure}

where $N_{tot} = N_0 + N_1 + N_2 + N_3$, $N_i=N_i(\vec r,t)$, $i=0,\ 1,\ 2,\ 3$ are the population  
number densities of the corresponding  
electron level; 
$N_{tot}$ is the total number of electrons participating in the  
 lasing process, 
$\gamma_{ij} \equiv 1 / \tau_{ij}$ are the transition  
rates from level $i$ to $j$, and $\gamma_{nr}$ is the non-radiative 
decay rate of the laser level 2. $\gamma_P \equiv 1 / \tau_P  $ is the transition  
rate due to homogeneous, constant, external pumping. Further  
$n_{ph} \equiv N_{ph} / N_{tot}$ is the  
photon number density, normalized to $N_{tot}$. The four level system is
chosen, because the stationary state, the threshold is easily established, i.e. 
$\partial_t N_i = 0 $, hence the above system of equations can 
be solved for the population inversion $n_2=N_2/N_{tot}$ to yield $n_2 =\frac{\gamma_P}{\gamma_P  + \gamma_{nr} +  \gamma_{21}\left(n_{ph}
    +1\right)}$, where it was assumed that $\gamma_{32}$ and  $\gamma_{10}$ are 
large compared to any other decay rate.\\
In the last step the laser rate equations are coupled to the microscopic 
transport theory by identifying the growth term in the photon diffusion with 
corresponding growth term in the derived equation for the energy density 
correlation $\gamma_{21}n_2=D/{-\xi^2}$. The equality is ensured by
finding an approbate self-consistent imaginary part
of the scatterers dielectric function ${\rm Im}\,\epsilon_s$ for any particular 
light frequency $\omega$ and any position $Z$ within the slab geometry.\\

The here developed theory of random lasing includes the regular 
self-consistency of the Vollhard-W\"olfe type for the diffusion 
coefficient including the interference effects of emitted light intensity. The
self-consistent results for the imaginary part $\rm Im\, \epsilon_s$ 
of the dielectric coefficient of the laser active scatterers are derived by
coupling the mesoscopic transport  to semiclassical laser rate equations. The
single particle self-energy $\Sigma(\omega)$ entering the single particle
Green's function, as in  Eq. (\ref{SP_Green_function_field}), is approximated
as $\Sigma=n \cdot T$, where $n$ is the volume
filling fraction of scatterers in the host medium and $T$ is the T-matrix of
the Mie scatterer. These internal resonances are renormalized, but
    nevertheless they allow for low
and stable laser thresholds. In Fig. \ref{corr1} the calculated correlation length $\xi$ is given in units
of the scatterer radius $r_0$ and presented as
a function of the scattering mean free path $l_s$ of the random system. The
bullets on the curves correspond to several filling fractions.
We find, that the correlation length increases with the mean free paths and
again decreases with higher pump intensities. That behavior can be interpreted
as self-balancing of sample energy at the threshold. On the abscissa $l_s$ is
given in units  of the scatterer radius $r_0$.The different curves  correspond
to different strengths of the pump intensity P, given in units of transition
rate $\gamma_{21}$. The displayed symbols along one graph correspond to
numerical evaluations for different filling fractions of the 
zinc oxide scatterers. In particular, the points from left to right 
correspond to filling fractions of $60\%$, $50\%$, $45\%$, $40\%$, and $35\%$.
The correlation or coherence length is found to decrease with
increasing pump intensity. Further, the dependence on the
scattering mean free path is decreased for stronger pumping.
This is in agreement with recent experimental results, as discussed
in reference \cite{Redding1}. The correlation length $\xi$ at the surface of the sample, Fig. \ref{corr1},
is displayed in Fig. \ref{corr2} across the sample thickness for a volume filling fraction of the scatterer of
$40\%$ and a pump rate of $P=0.1\gamma_{²1}$ in the upper panel. This is accompanied by the calculated photon number density in the middle panel
as well as the scattering mean free path $l_s$ as a function of the 
sample depth $Z$. The self-consistent scattering mean-free path $l_s$ Fig. \ref{fig04} is
displayed for various filling fractions $35\%$..$60\%$. The density dependency
can be clearly observed, but additionally and even more important the variation of the scattering mean free path $l_s$ across
the width of the sample is especially noteworthy. It clearly shows the
influence of lossy boundaries on the inversion and therefore the
self-consistent gain and gain saturation depending on the position. Near the
surface the mean free path is in units of $r_0$ increasing. However it
increases counter-intuitively stronger for low filling fractions. Consequently
the approach of some ballistic limes in low filling raises expectations that
the lossy boundary will gain importance. This is an important result which has
not been reported before.

\section{Conclusion}
In summary, we considered a system of randomly positioned laser active ZnO
scatterers and we developed a systematic theory for 
random lasing in finite sized disordered systems. 
The self-consistent theory of light localization, including interference 
effects, is coupled to the laser rate equation in order to obtain the corresponding nonlinear
gain of the lasing system. The calculated averaged correlation length of the occurring lasing spots is systematically studied and 
found to exhibit a characteristic dependence on the scattering
mean free path of the sample in dependency to the in-depth position. Further
the self-consistent scattering mean free path $l_s$ depends
on the position with respect to lossy boundaries. Therefore we conclude that the complex valued T-matrix "off-shell" of the single scatterer in self-consistent stationary state
significantly varies across the sample and therefore the scattering strength of
the particle is driven by so-called electromagnetically induced transparency
processes (EIT) which obviously occur not only in high non-equilibrium of the
ultrafast regime, but they do also occur due the amount of coherent intensity
present in the scattering sample. The latter is for homogeneous pumping
higher in the samples center.


\bibliography{foo}
	\bibliographystyle{plain}

\end{document}